\documentclass[
 reprint,
 amsmath,amssymb,
 aps,
 prf
]{revtex4-1}

\usepackage{graphicx}
\usepackage{dcolumn}
\usepackage{bm}

\usepackage{amssymb}
\usepackage{amsmath}
\usepackage{mathtools}

\begin{document}


\title{Equilibrium and stability of two-dimensional pinned drops}

\author{J. Gra\~na Otero}
\email{jose.grana@uky.edu}
\affiliation{279 RGAN. Dept. of Mechanical Engineering. University of Kentucky, KY, USA}
\author{I. E. Parra Fabi\'an}
\affiliation{Center for Computational Simulation (CCS). Universidad Polit\'ecnica de Madrid. Espa\~na}

\date{\today}

\begin{abstract}
Superhydrophobicity relies on the stability of drops's interfaces pinned on sharp edges to sustain non-wetting (Cassie-Baxter) equilibrium states. Gibbs already pointed out that equilibrium is possible as long as the pinning angle at the edge falls between the equilibrium contact angles corresponding to the flanks of the edge. However, the lack of stability can restrict further the realizable equilibrium configurations. To find these limits we analyze here the equilibrium and stability of two-dimensional drops bounded by interfaces pinned on mathematically sharp edges. We are specifically interested on how the drop's stability depends on its size, which is measured with the Bond number $Bo = (\mathcal{W}_d/\ell_c)^2$, defined as the ratio of the drop's characteristic length scale $\mathcal{W}_d$ to the capillary length $\ell_c = \sqrt{\sigma/\rho g}$. Drops with a fixed volume become more stable as they shrink in size. On the contrary, open drops, i.e. capable of exchanging mass with a reservoir, are less stable as their associated Bond number decreases.
\end{abstract}


\maketitle

\section{Equilibrium of drops.}

\begin{figure}[h!]
	\centering \includegraphics[width=\linewidth]{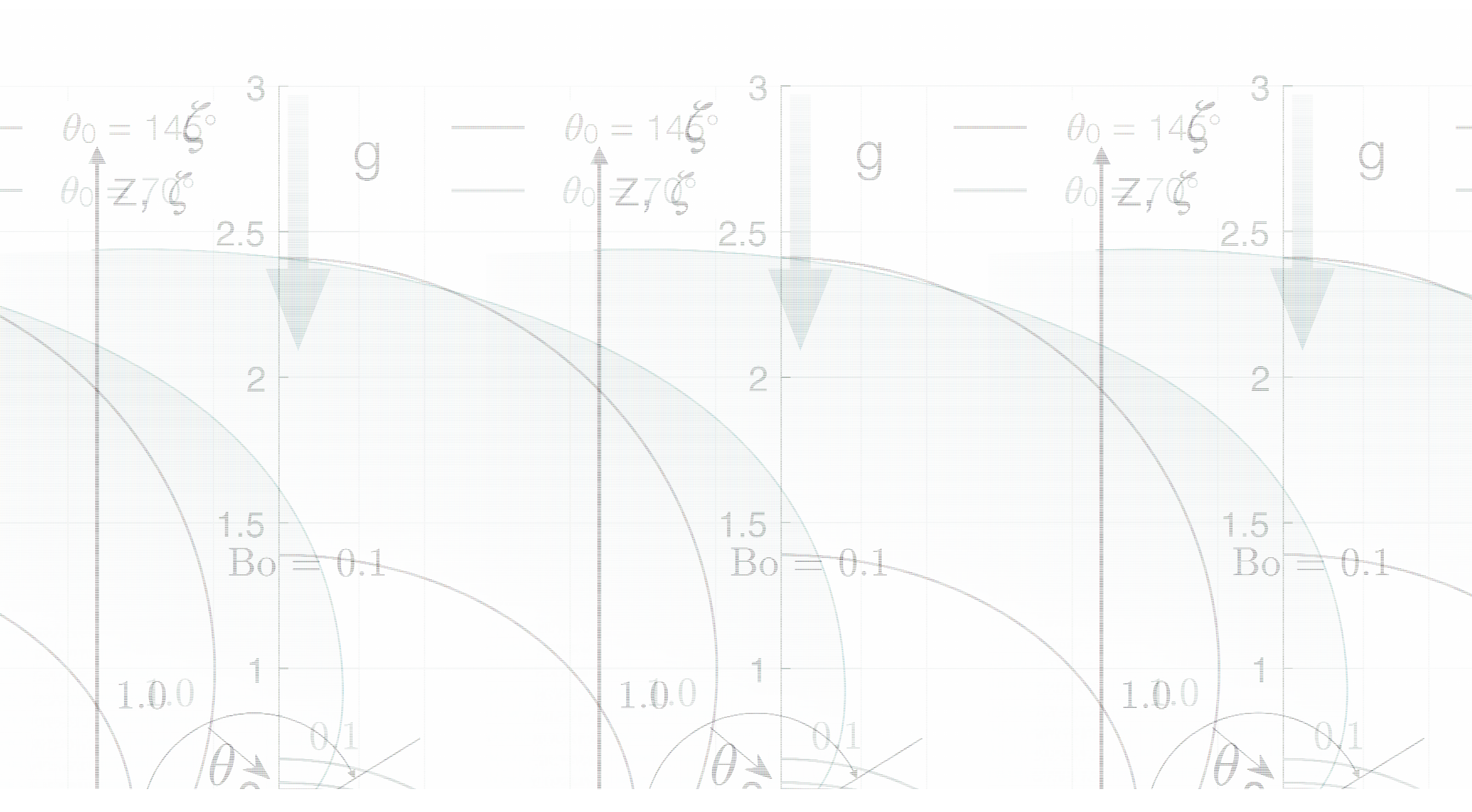}
	\caption{At the left, sketch of a drop at equilibrium under gravity pinned symmetrically on mathematically sharp edges. The details of the solid substrate (if any) between the edges has been deliberately left undefined because it does not play an essential role in the equilibrium. This configuration thus includes cases of constant volume when the liquid somehow limited either by a solid surface or by another interface, and of constant pressure drops (see details below). At the right, drop's interface  shapes for representative values of the Bond number $Bo = (\mathcal{W}_d/\ell_c)^2$, and pinning contact angle $\theta_0$.}
	\label{Intro}
\end{figure}

We study here the equilibrium and stability of drops resting symmetrically pinned on two sharp edges as sketched in Fig.~\ref{Intro}. The horizontal separation between the edges $2\mathcal{W}_d$ is a measure of the drop's size, to be compared with the capillary length $\ell_c = \sqrt{\sigma/(\rho g)}$, with $\sigma$ and $\rho$ the liquid's surface tension and density respectively and $g$ the acceleration of gravity. Their ratio defines the Bond number of the problem $Bo = \rho g \mathcal{W}_d^2/\sigma = (\mathcal{W}_d/\ell_c)^2$, which measures the relative importance of gravity compared with surface tension forces at the scale $\mathcal{W}_d$ of the drop.


The limit of large Bond numbers corresponds to a drop extending over lengths $\mathcal{W}_d \gg \ell_c$ much larger than the capillary length. This is the case of everyday life, when the dominant gravity squeezes the free surface flat over most of the drop except near the edges where surface tension becomes significant and bends the flat interface as required by pinning.

The limit of small Bond numbers on the other hand corresponds to a separation between edges $\mathcal{W}_d \ll \ell_c$ much smaller than the capillary length, so gravity effects are weak at this scale. Drops with these sizes tend to be circular (or spherical in 3D problems) on earth where capillary lengths are typically of the order of millimeters and viscous effects damp quickly long-wave fluctuations. This is typically the scale of surface micro-patterning features used to promote superhydrophobicity, either natural or artificially fabricated\cite{Plants, BHUSHAN2011}, which has length scales of the order of microns or smaller. The liquid interfaces pinned between these features fall thus in this limit. 

On the other hand, this particular case of drops resting on superhydophobic substrates is an example of a class of problems involving a wide range of lengths scales. Drop's sizes are  typically in these cases, as for instance rain drops, of the order of the capillary length, therefore with associated Bond numbers of order unity. However, as mentioned, the interface underneath is pinned on the micro-features and therefore corrugated at this much smaller scale compared with the drop's scale. In these cases, the wide length contrast can be advantageously exploited to gain insight in the otherwise highly complex problem of equilibrium and stability of these configurations. For instance, it can be shown that the interfaces at the drop's scale and at the micro-scale can be calculated independently of each other and then, using matching conditions, make them part of the same volume of liquid, i.e. drop. 

Here we focus on the important problem of the stability of equilibrium configurations. We are in particular interested in understanding how the length scale of the drops affects its stability. These analysis are often complicated\cite{Lenz2000}. Taking advantage of the variational formulation available in this problem, we use instead turning-point arguments\cite{PoincareII, Maddocks,Steen} which are more straightforward in this case.


\subsection{Formulation}

The problem of finding the drops' equilibrium shapes and their stability can be formulated as the minimization of the energy functional:

\begin{gather}
	\label{EnergyFunctional}
	\mathcal{E} = 
	\int_0^{s_a}{\left(\sqrt{\dot{\xi}^2 + \dot{\zeta}^2} + (P_0\zeta - (1/2)Bo\zeta^2)\dot{\xi} \right) ds}
\end{gather}
where $\mathcal{E}$ is the energy made dimensionless with $\sigma \mathcal{W}_d$ and $(\xi, \zeta)$ are, respectively, the $x$ and $z$ coordinates of the interface made dimensionless with the length $\mathcal{W}_d$. To accommodate the largest class possible of shapes, the interface is parametrized with the arc-length $s$, also made dimensionless with $\mathcal{W}_d$, and the dot represents the $s$-derivative. The upper limit $s_a$ of the line integral corresponds to the axis of symmetry $\xi = 0$, so $s_a$ the interface's half-length. 

\begin{figure*}
	\centering \includegraphics[width=.45\linewidth]{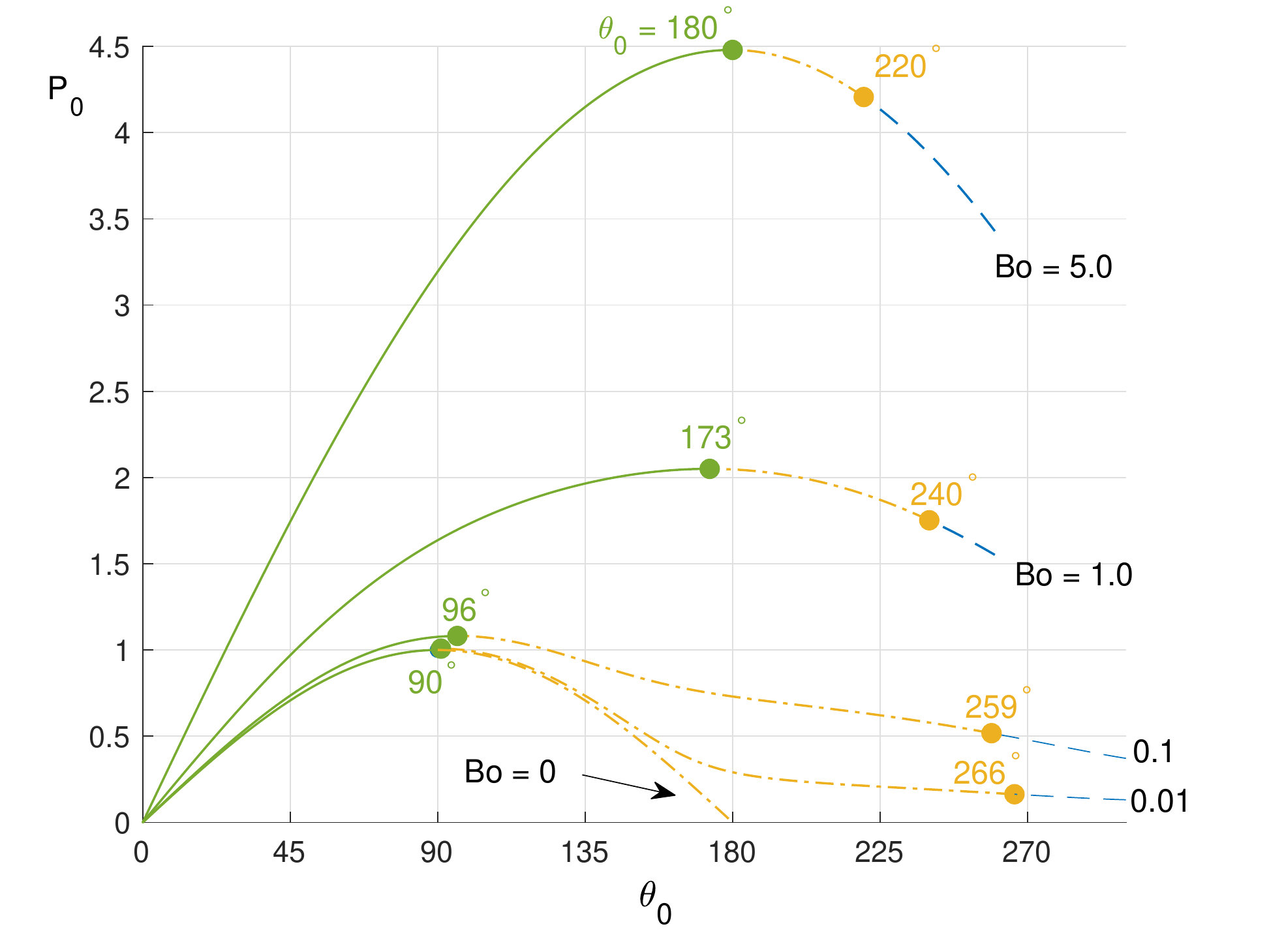} 
	\includegraphics[width=.45\linewidth]{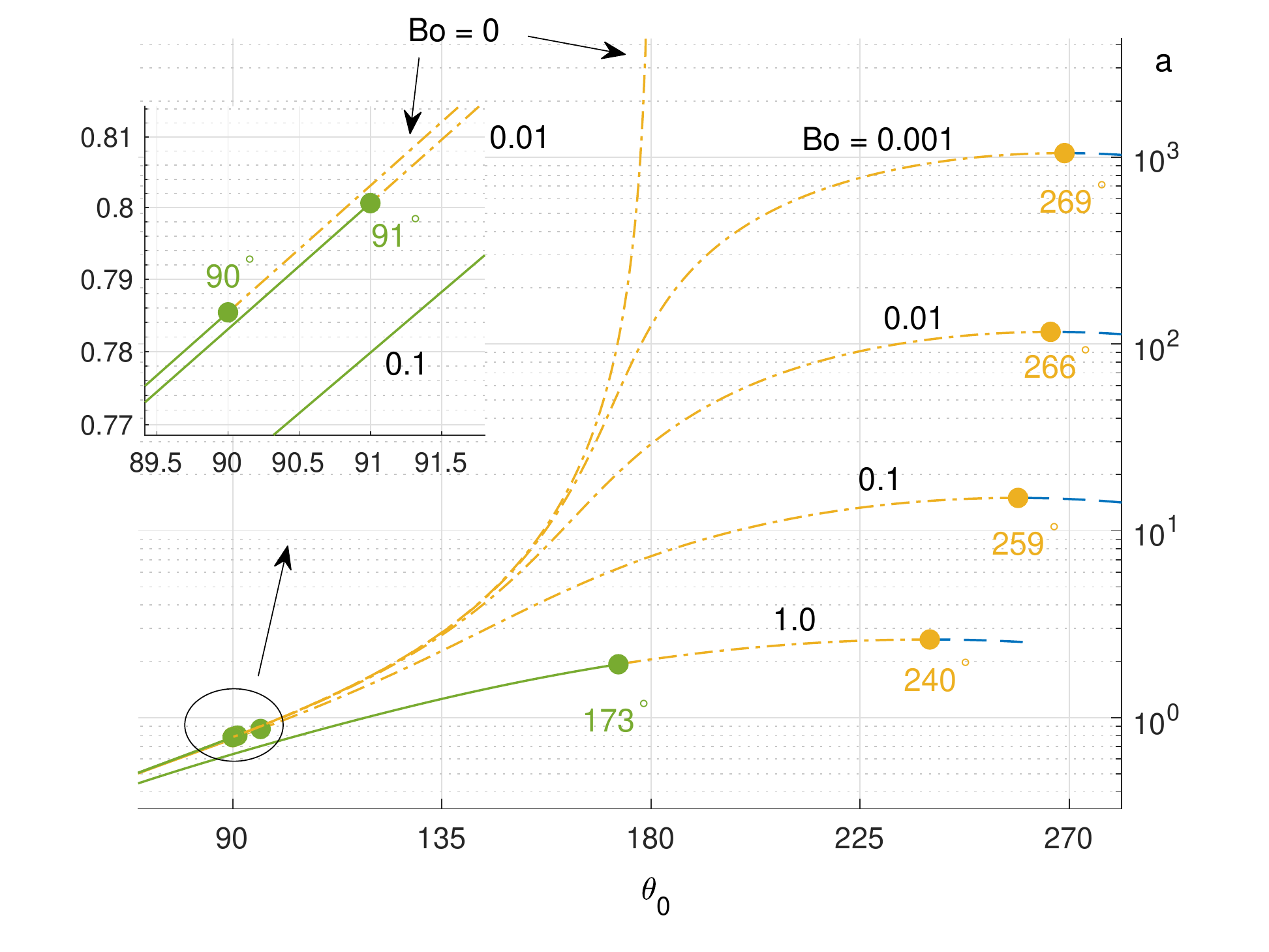}
	\caption{The pressure eigenvalue $P_0$ at the left and the drop's area $a$ at the right as functions of the pinning angle $\theta_0$ for several values of the Bond number $Bo = (\mathcal{W}_d/\ell_c)^2$. The green dots mark the pressure eigenvalue $P_0$ maxima where the solutions change from constant-pressure (solid green lines) to constant-volume stable (dash-dot lines), whereas the equilibrium become unstable past the yellow dots, i.e. the maxima of the drop's area $a$. Notice the divergent area at $\theta_0 = 180^\circ{}$ and the lack of solutions beyond this limit angle in the case of zero Bond number, i.e. in the absence of gravity. Notice that the invariance of the problem \eqref{ProblemParametricEigenvalueDrop} under reflections about the horizontal axis translates in the same symmetry in this diagram, so the critical receding angles are the opposite of those shown.}
	\label{EigenvalueAreaPlanar}
\end{figure*}

The pressure inside the drop, measured with respect to the ambient pressure $p_\infty$ and made dimensionless with $\sigma/\mathcal{W}_d$, is $P_0 - Bo~\zeta$. The constant $P_0 = (p(z= 0) - p_\infty)\mathcal{W}_d/\sigma$, i.e. the pressure at zero height is an eigenvalue (the pressure eigenvalue hereafter) of the problem and must be determined as part of the solution. It represents the value of the pressure required to produce a specific value of the pinning angle $\theta_0$ at the edge. From a practical point of view, this is for instance the pressure that should be used to make the drop grow or shrink by injecting or removing fluid though a tube with the outlet at $z = 0$.

The Euler-Lagrange equations for the extremals of the functional \eqref{EnergyFunctional} are:
\begin{subequations}
\label{ProblemParametricEigenvalueDrop}
	\begin{gather}
		\label{DiffEquationsXi}	
		\ddot{\xi} = - \dot{\zeta} \, (P_0 - Bo \, \zeta) \\
		\label{DiffEquationsZeta}
		\ddot{\zeta} = \dot{\xi} \, (P_0 - Bo \, \zeta)
	\end{gather}
	to be integrated with boundary conditions:
	\begin{gather}
		\label{PinningBC}
		\xi(s = 0) = 1, \quad \zeta(s = 0) = 0 \\
		\label{AngleBC}
		\dot{\xi}(0) = -\cos(\theta_0), \quad \dot{\zeta}(0) = \sin(\theta_0) \\
		\label{AxisSymmetryBC}
		\xi(s = s_a) = \dot{\zeta}(s = s_a) = 0 
	\end{gather}
which enforce: i) pinning of the drop at the edge $\xi = 0$, $\zeta = 0$ (\ref{PinningBC}), ii) the angle $\theta_0$ of the interface at the edge \eqref{AngleBC}, and iii) symmetry of the drop with respect to the vertical axis $\xi = 0$. Angles are measured through the drop with respect to the horizontal, as shown in Fig.~\ref{Intro}. Notice that despite the system being of fourth order, the problem contains the eigenvalue $P_0(Bo, \theta_0)$ as an additional unknown, requiring thus five boundary conditions. Notice also that the two boundary conditions \eqref{AngleBC} enforcing the angle $\theta_0$ are not independent since they are actually linked by the condition $\dot{\xi}^2(0) + \dot{\zeta}^2(0) = 1$ resulting from $s$ being the arc-length parameter. 

The choice of the pinning angle as independent parameter is motivated by the results that show that the pressure eigenvalue turns out to be a one-to-one function of the pinning angle for each Bond number as Figure \ref{EigenvalueAreaPlanar} shows.

Additionally, the drop's area $A$, or its scaled version $a(Bo, \theta_0) = A/2\mathcal{W}_d^2$, (where the inconsequential factor 2 in the denominator is introduced for convenience) can be obtained as well from the solution as 
	 \begin{gather}
		\label{DropArea}
		a = A/2\mathcal{W}_d^2 = -\int_0^{s_a}{\zeta d\xi}
	\end{gather}
\end{subequations}
\begin{figure}
		\centering \includegraphics[width=\linewidth]{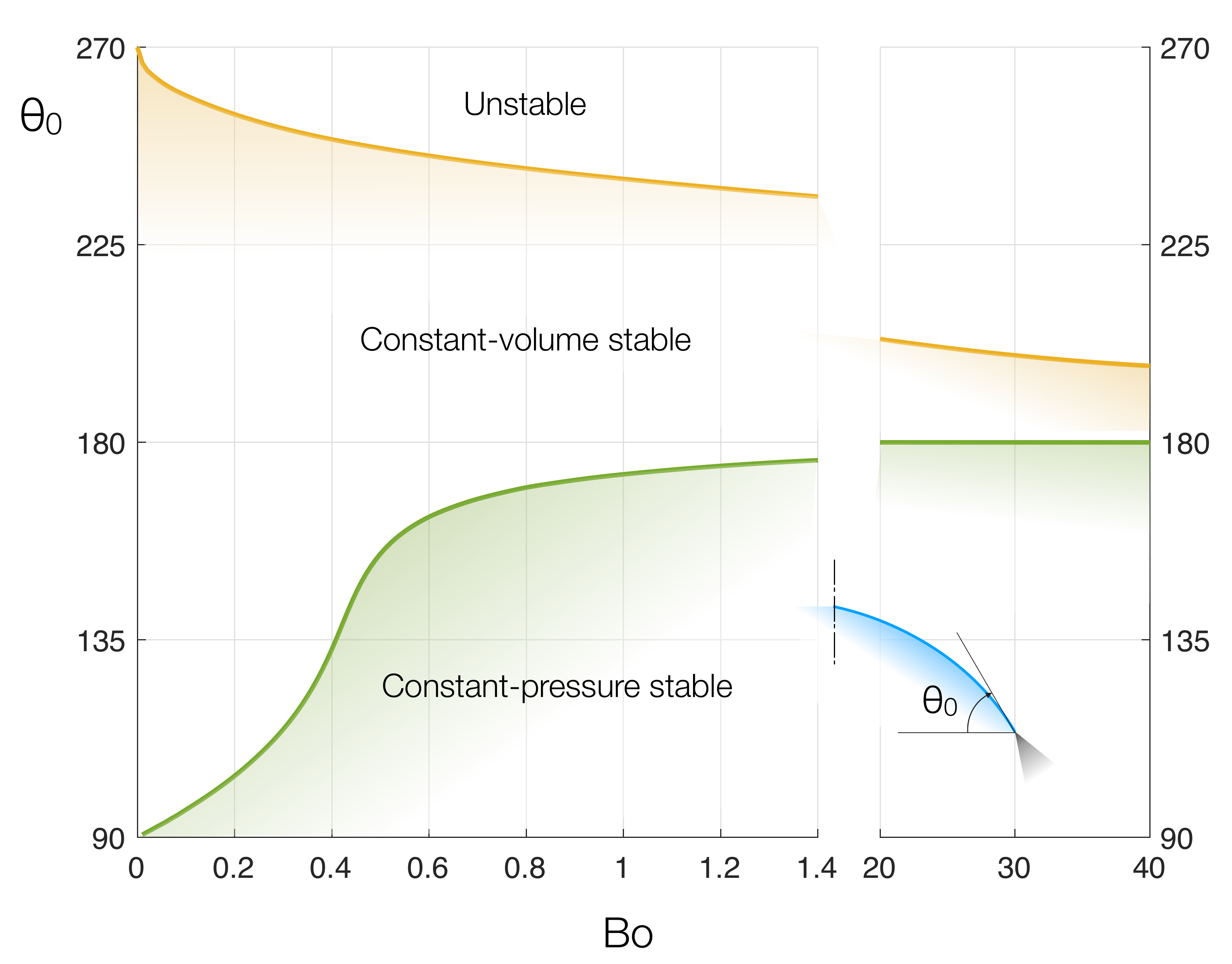}
	\caption{Stability map for finite planar symmetrical drops. The limits of stability are represented by the solid lines, which give the critical pinning angle $\theta_0$ of the loss of stability of drops subject to constant-pressure (green line) or to constant-volume (yellow line) perturbations. These limits of stability correspond\cite{Maddocks} to maxima of the curves $P_0(a)$ (green dots and line in Figure \ref{EigenvalueAreaPlanar}) in the case of the constant-pressure limit, and to maxima (yellow dots and lines in Figure \ref{EigenvalueAreaPlanar}) of the curves $a(P_0)$ in the constant-volume case. Drops in the region below the green line are thus stable, those between the yellow and green lines are unstable to constant-pressure perturbations but stable to constant-volume perturbation, and finally those above the yellow line are unstable. Notice that the invariance of the problem \eqref{ProblemParametricEigenvalueDrop} under reflections about the horizontal axis translates in the same symmetry in this diagram, so the critical receding angles are the opposite of those shown. The result $\theta_0 = 90^{\circ{}}$ for $Bo = 0$ was previously studied theoretically with other methods by Speth and Lauga\cite{Speth_2009} and experimentally by Gau et al's\cite{Gau46}.}
	\label{StabilityMap}
\end{figure}

\subsection{Drops} Fig.~\ref{EigenvalueAreaPlanar} shows the pressure eigenvalue $P_0$ and the area $a$ of the drop as functions of the pinning angle $\theta_0$ at the edge for several values of the Bond number $Bo$. It can be noticed that there are solutions for all values of the angle $\theta_0$ in the range $0^\circ{} \le \theta_0 \le 360^\circ{}$ for non-zero Bond numbers. However, in the case zero Bond number $Bo = 0$, the absence of gravity limits the existence of solutions to angles in $0^\circ{} \le \theta_0 \le 180^\circ{}$. The reason can be easily understood considering that without gravity only circular shapes are possible, with the radius $R = \mathcal{W}_d/\sin(\theta_0)$ and the drop's area $4a = (2\theta_0 - \sin(2\theta_0))/\sin^2\theta_0$, which diverge as $\theta_0 \rightarrow 180^\circ{}$. Or in physical terms, only gravity is capable of deforming the drop below the pinning corner ($\theta_0 > 180^\circ{}$), so the circular interface, i.e. $Bo$ exactly zero, is incompatible with angles $\theta_0$ larger than 180$^\circ{}$.

The problem \eqref{ProblemParametricEigenvalueDrop} is invariant under reflections about the horizontal axis $\xi$, that is to say under the transformation $\zeta \rightarrow -\zeta$, $\theta_0 \rightarrow -\theta_0$, $P_0 \rightarrow -P_0$, $a \rightarrow -a$ (negative areas are to be understood below the horizontal line as suggested by \eqref{DropArea}. Thus, the functions $P_0(\theta_0)$ and $a(\theta_0)$ in Fig.~\ref{EigenvalueAreaPlanar} can be extended to negative angles by changing the sign of $P_0$ and $a$, whereas the stability map of Figure \ref{StabilityMap} is symmetrical with respect to the horizontal axis.

\begin{figure}[h!]
	\centering \includegraphics[width=0.9\linewidth]{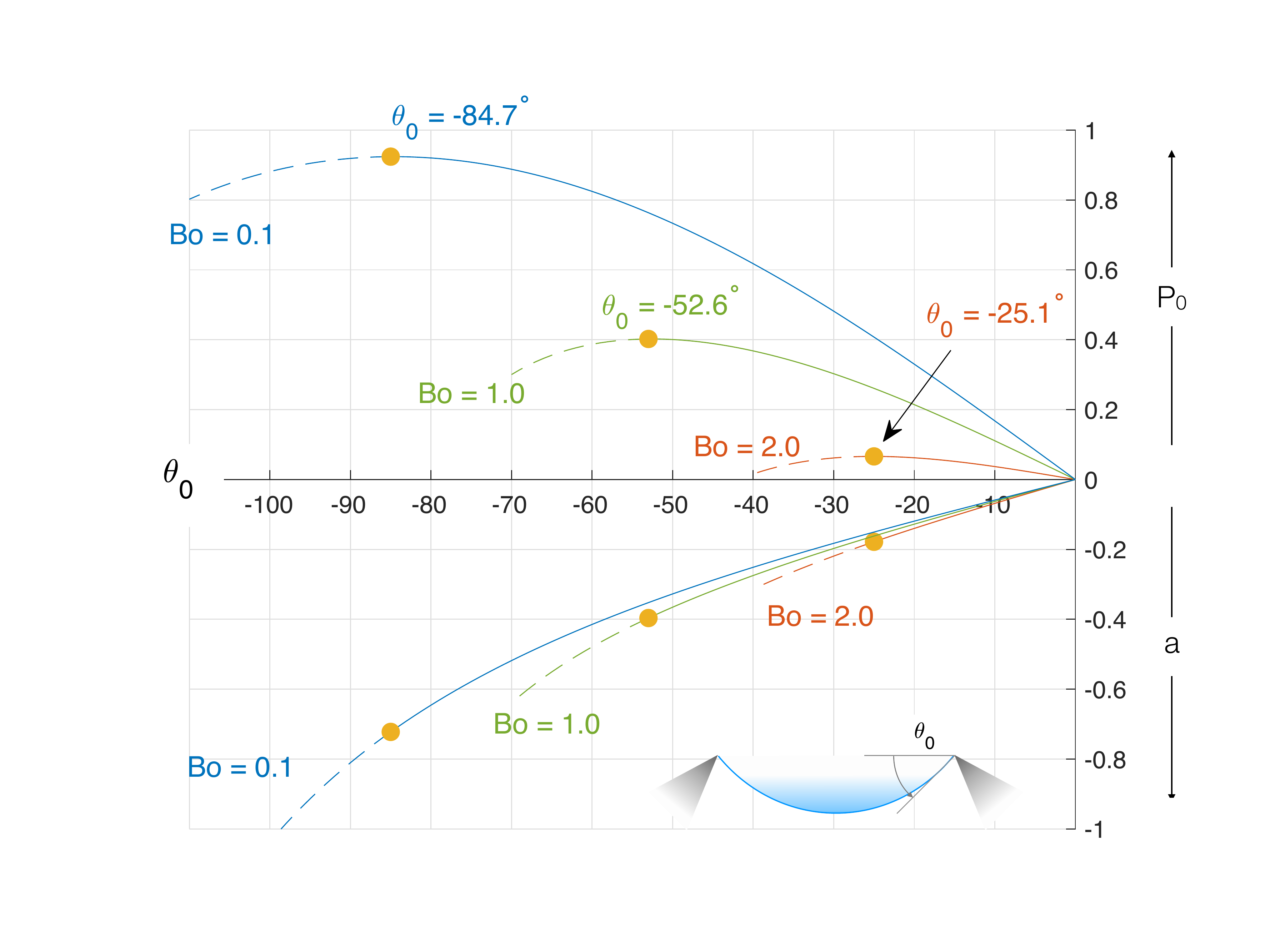}
	\centering \includegraphics[width=0.8\linewidth]{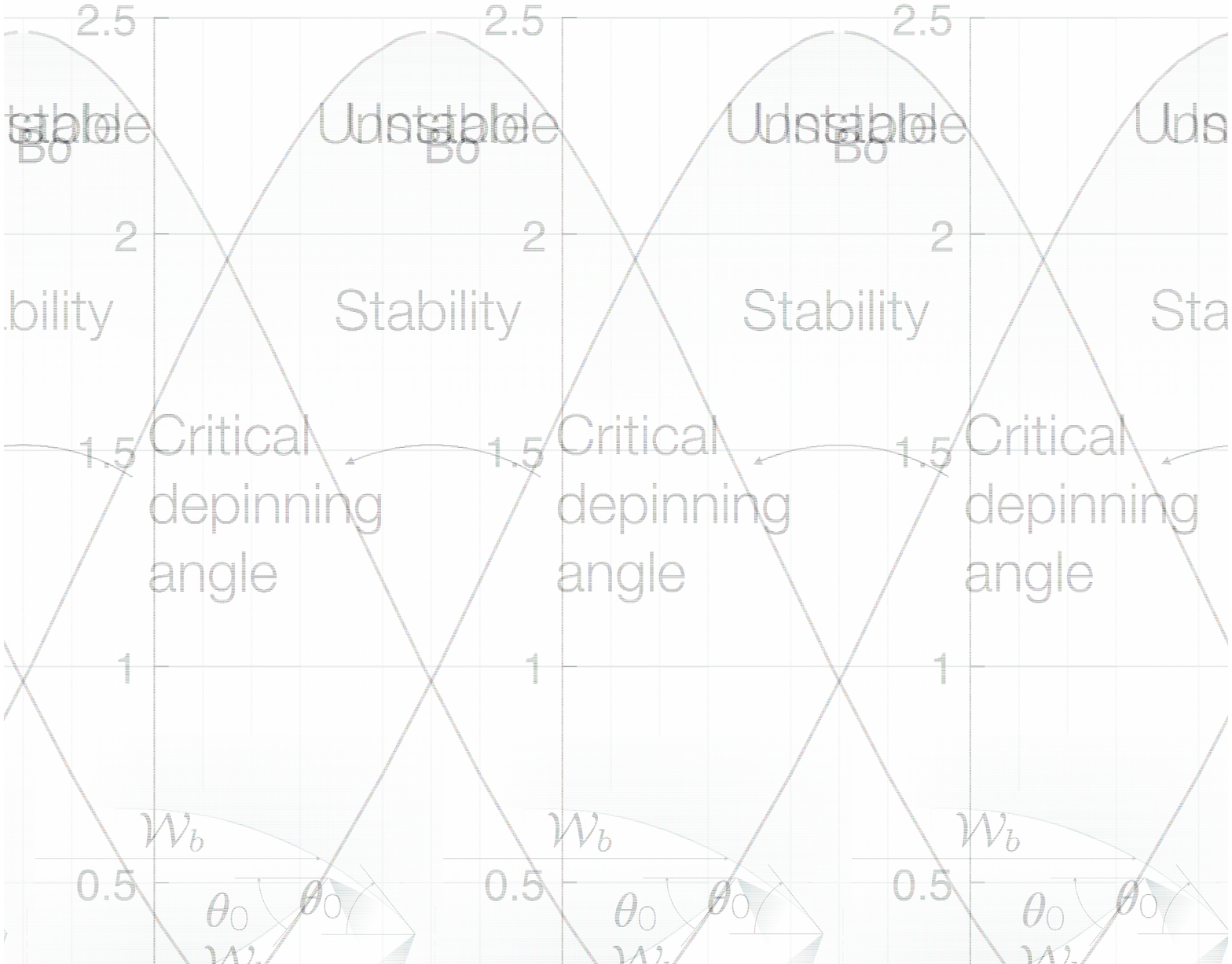}
	\caption{At the top, for representative values of the Bond number $Bo = (\mathcal{W}_d/\ell_c)^2$, the pressure eigenvalue $P_0$ and the area $a$ as functions of the pinning angle $\theta_0$ (negative below the horizontal) for interfaces supporting the liquid, that is with the liquid above the interface. As shown in the inset, the interface is below the horizontal has corresponding negative pinning angles and areas in accordance with the convention defined by \eqref{DropArea}. The solid (dash) lines represent the stable (unstable) equilibrium branches, whereas the yellow dots, labelled with the critical depinning angles, mark the folds (maxima of $P_0(\theta_0)$) where the change of stability occurs. At the bottom, the constant-pressure stability limits for hanging drops. The two insets display the shape of the critical interface shapes. Notice the anticipated symmetry of the problem, i.e. $\theta_0 \rightarrow -\theta_0$, $P_0 \rightarrow -P_0$, $a \rightarrow -a$.}
	\label{BelliesP0a}
\end{figure}

\subsection{Inverted drops}

Configurations as those in Fig.~\ref{BelliesP0a}, i.e. with the liquid above the interface, can be obtained with the same formulation but reversing the sign of the curvature, which physically changes the liquid's side with respect to that of the previous section. Proceeding in much the same way as before one can obtain the eigenvalue $P_0$ and the area $a$ as functions of the pinning angle $\theta_0$ and of the Bond number as shown in Fig.~\ref{BelliesP0a}. The changes of stability correspond to maxima of the pressure eigenvalue $P_0(\theta_0)$, so plotting the critical angles for each Bond number gives finally the stability diagram of Fig.~\ref{BelliesP0a}. Notice that both positive (interface above the horizontal) and negative pinning angles are possible. However, as mentioned, the symmetry with respect to the horizontal makes unnecessary to include both. We have though included the complete range of angles $\theta_0$ to ease the interpretation.

%

\section{Stability}

Some of the equilibrium solutions of the problem \eqref{ProblemParametricEigenvalueDrop} may actually not be realizable if they fail to be stable. As we show below, this stability analysis turns out to be related to the stability of non-wetting states on superhydrophobic substrates, which is a major concern in applications and has triggered extensive experimental work. Caution needs to be exerted however, because this literature is not homogeneous and sometimes it even appears contradictory. The reason is that experimental conditions often enforce subtle constraints on the admissible motions the drops can undergo, with the consequent impact on the stability.

\subsection{Stability types.} Two main classes of perturbations can be typically found depending on whether the volume of the drop is kept constant or not \cite{Steen}. In the case of unconstrained, i.e. constant-pressure perturbations, the volume of drop is left free to fluctuate when a given equilibrium solution (and therefore for a specific fixed value of the pressure eigenvalue $P_0$) is perturbed. This is the relevant type of stability in cases where the drop is somehow open to a reservoir which can accommodate its volume changes. This is the case for instance in experiments where the liquid is supplied to the drop through some form of tubing as in Oliver et al.'s experiments\cite{HuhMason}. They considered the equilibrium and stability of axisymmetric drops pinned at the sharp-edged rim of a flat horizontal circular surface with a drill at the center used to supply the liquid. The pressure is set externally to gradually adjust the drop's size, and liquid can flow in and out of the drop when its shape fluctuates. Constant pressure perturbations are also appropriate in a class of experiments where a hydrophobic substrate is subject to a pressurized liquid enclosed in a large chamber\cite{SuperHBreakdownIII, micropillars, Hensel:2014aa}. 



On the other hand, the constant-volume stability is appropriate in experiments where the drop is closed in the thermodynamic sense, i.e. not able to exchange mass with the surroundings. This is the case for instance when a drop is \emph{released}, i.e. detached from the generating device, and on top of a hydrophobic substrate\cite{Reentrant, Haimov:2013aa}; impacting drops belong as well in this class\cite{DropBouncing,DropImpactI,DropImpactMD}. Or in experiments with evaporating or condensing but otherwise closed drops resting on a micropatterned substrate\cite{XRays}. The volume can be assumed constant in these cases because, typically, the time scale for drop's size growth or shrinkage due to these slow mass transfer processes is much larger than the time of relaxation back to rest (or of destabilization) if the equilibrium is perturbed.

\subsection{Stability map.} We will confine the stability analysis to planar pinned perturbations, so the perturbed solutions are still planar and pinned and the energy functional of \eqref{EnergyFunctional} is appropriate. This excludes therefore longitudinal Rayleigh-Plateau-like instabilities as those already analyzed theoretically for instance by Langbein\cite{langbein_1990} and more recently experimentally by Herminghaus et al.\cite{HerminghausMaterials}. However, typically, drops are usually small enough to be well within the stability region of this type of instability which is thus usually irrelevant in superhydrophobicity. 

We are neither interested in the destabilization dynamics so we can just check which of the equilibrium extremals correspond to local minima of the energy functional \cite{Guelfand}, i.e. to solutions with a positive definite $\mathcal{E}$'s second Fr\'echet-derivative. We exploit thus the variational structure of the problem and use turning-point methods, first introduced by Poincar\'e \cite{PoincareII} and discussed more recently by Maddock's \cite{Maddocks} (see also \cite{Steen}). In short, the stability can be studied in the preferred plane $P_0 - (-\mathcal{E}_{P_0})$, where $P_0$ is the eigenvalue of the problem and $\mathcal{E}_{P_0} = -a$ is the partial derivative of the functional $\mathcal{E}$ with respect to $P_0$. Translating his results to our problem, it turns out that unconstrained extremals change stability at folds of the solution branches in the $P_0-a$ plane, that is at local extrema of the pressure eigenvalue function $P_0(a)$; whereas constant volume extremals change stability at local extrema of the drop's area $a(P_0)$. He further proved that constant-volume stable equilibria are also constant-pressure (unconstrained) stable, but not the reciprocal. Figs.~\ref{EigenvalueAreaPlanar} and \ref{BelliesP0a} show as green and yellow dots those points, and in solid green, dash-dot yellow and dash blue unconditionally stable, constant-volume stable and unstable solutions respectively. From these plots, the stability map in the $Bo-\theta_0$ plane of Fig.~\ref{EigenvalueAreaPlanar} is derived, showing the region of drops stable under constant-pressure perturbations (below the green line), stable under constant-volume perturbations (between the yellow and green lines) and unstable drops. It can be noticed the common limit $\theta_0 = 180^\circ{}$, attained at large Bond numbers, of both limits of stability. This limit corresponds to a semi-infinite liquid layer for which constant-pressure and constant-volume perturbations are clearly equivalent.

\section{Discussion}

We have seen that drops subject to gravity can be found in static equilibrium supported interfaces attached to sharp edges. Contrary to smooth surfaces where the contact angle is well defined, sharp edges are singular (in the mathematical sense) because the normal vector is not well defined on them, so the angle of the attached (pinned) interface is as well undetermined. Equilibrium provides only relationships of the kind of those in represented in Figures \ref{EigenvalueAreaPlanar} and \ref{BelliesP0a}, giving the pressure or the area that drops should have for a given value of the pinning angle. Equilibrium is thus possible in broad ranges of pinning angles. However, as shown, the stability of these configurations restricts the realizable solutions to narrower domains with bounds which depend on the size of the system through the Bond number, the ratio of the characteristic length of the problem to the capillary length. 



%
\begin{figure}
	\centering \includegraphics[width=.5\linewidth]{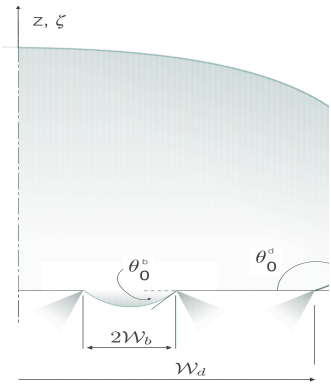}
	\caption{A sketch of a multiscaled drop model, with only two scales $\mathcal{W}_b \ll \mathcal{W}_d$, associated with a macro-interface pinned on edges separated a distance $2\mathcal{W}_d$ and the micro-interface of a small belly pinched between $2\mathcal{W}_d$-wide edges. Notice that the microscale belly is an inverted drop as in Figure \ref{BelliesP0a} with negative $\theta_0$. As indicated, $\theta_0^d$ and $\theta_0^b$ are the respective pinning angles of the top drop's interface and of the belly underneath. The corresponding pressure eigenvalues $P_0^i = (p(z=0) - p_\infty)\mathcal{W}_i/\sigma$ are linked by the same pressure $p(z=0)$  at zero height, i.e. the horizontal through the edges.}
	\label{Multiscale}
\end{figure}

We show next how these results and ideas can be used to calculate multiscale interfaces' shapes as those usually adopted by drops at equilibrium on superhydrophobic substrates.

\subsection{Multiscale drops.} 


Consider the configuration sketched in Fig.~\ref{Multiscale}, with two sharp edges separated a distance $2\mathcal{W}_d$ supporting symmetrically a planar drop which, in addition, has a section of its bottom interface pinched between two other edges separated a distance $2\mathcal{W}_b$. In the limit $\mathcal{W}_b \ll \mathcal{W}_d$, this represents a model of a drop in equilibrium on a micropatterned superhydrophobic substrate. Despite the simplicity of the model, the discussion below is qualitatively applicable to 3D geometries and helps explaining the observed phenomenology. On the other hand, it can be easily generalized to more realistic configurations with multiple edges underneath the drop, and even with a distribution of sizes. 

\begin{figure}
	\centering \includegraphics[width=.8\linewidth]{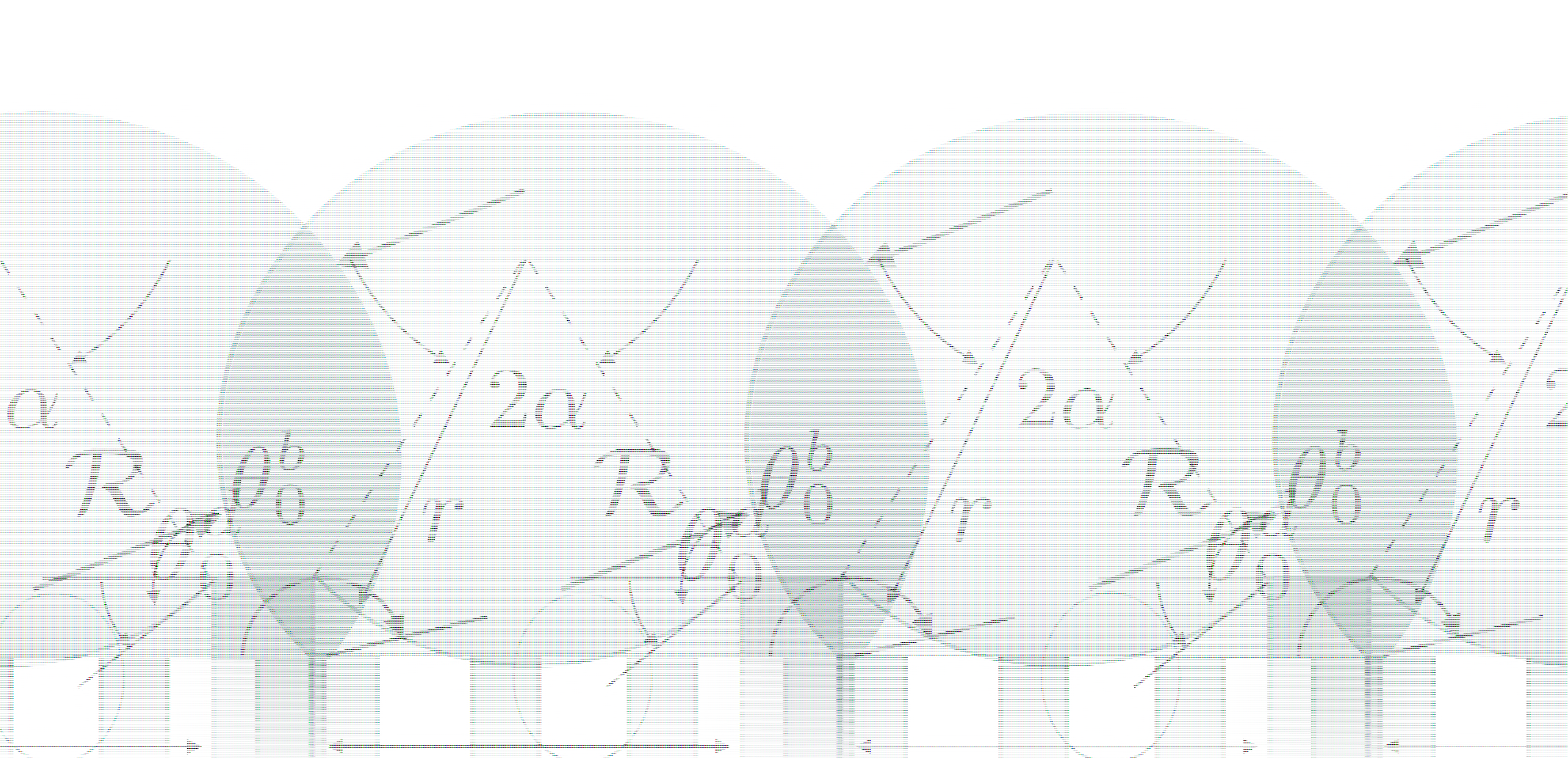}
	\caption{Geometry of the interface pinned between features of a model superhydrophobic substrate with a characteristic micro-scale length $\mathcal{W}^b$. The radius of curvature $r$ of the micro-interface is related to the angle $\alpha = \theta_0^b$ as shown.}
	\label{Underneath}
\end{figure}

The drop's interface can be considered as composed of disconnected interfaces, each one pinned between different edges. Each one can be calculated separately as shown above, defined by their corresponding pressure eigenvalue and pinning angle. However, since they belong to the same volume of liquid, they must have the same pressure distribution and therefore the same value $p(z=0)$:
\begin{gather}
	\label{Compatibility}
	P_0^b(\theta_0^b) = P_0^d(\theta_0^d)\sqrt{\frac{Bo^b}{Bo^d}}
\end{gather}
This is thus the condition that determines each interface's pinning angle $\theta_0^b$ and $\theta_0^d$, or more precisely, the relationship between them, in terms of their respective Bond numbers $Bo^b = (\mathcal{W}_b/\ell_c)^2$, $Bo^d = (\mathcal{W}_d/\ell_c)^2$.

Consider for instance the case when the scales are widely different $\mathcal{W}^b \ll \mathcal{W}^d$, so $Bo^b \ll Bo^d$ as well. According to \eqref{Compatibility}, the pressure eigenvalue $P_0^b$ for the micro-interface underneath is much smaller than $P_0^d$, the pressure eigenvalue associated with the macro-interface. As a result, as Fig.~\ref{BelliesP0a} shows, the micro-interface must have a close-to-zero pinning angle $\theta^b_0 \approx 0^\circ{}$, and therefore an almost flat shape, as has clearly been observed experimentally for instance by Haimov et al.\cite{Haimov:2013aa} (see also Schellenberger et al.\cite{MaxPlank}). However, as the drop's size decreases to values not much larger than the scale of the substrate microroughness, so the the factor $\sqrt{Bo^b/Bo^d}$ is not much smaller than unity anymore, then the pressure eigenvalues $P_0^d$ and $P_0^b$ become comparable and accordingly the pinning angle $|\theta_0^b|$ associated with the interface at the microstructure increase to produce bellies with significant curvature.

Physically this can be understood as follows. Drops with sizes of order $\mathcal{W}_d$ induce pressure jumps across their interface of order $\sigma/\mathcal{W}_d$ (a factor 2 should be included in the case of a two-dimensional interface). The pressure at zero height is thus of order $p(z = 0) - p_\infty \approx \rho g \mathcal{W}_d + \sigma/\mathcal{W}_d$. On the other hand, the pressure jump across the interface underneath, pinned between edges separated a distance $\mathcal{W}_b \ll \mathcal{W}_d$, is of order $\sin(\theta_0^b)\sigma/\mathcal{W}_b$, which is much larger than $\sigma/\mathcal{W}_d$ unless $\theta_0^b$ is much smaller than unity (see Figure \ref{Underneath}). Thus, the only way to match the pressure distributions due to each interface is by having $\sin(\theta_0^b) \approx \theta_0^b \sim \mathcal{W}_b/\mathcal{W}_d$, i.e. an almost flat interface underneath between the microstructure. However, when the drop scale $\mathcal{W}_d$ decreases to sizes comparable to those of the superhydrophobic microstructure, both interfaces must have comparable curvatures, i.e. $\theta_0^b$ is not small anymore. 

This is in fact, the basic mechanism of superhydrophobicity: the scale of  micropatterning must be much smaller than the size of the drops it should support. In that case, the interface underneath in contact with the microstructure is approximately flat. However, as the size of the drop and of the microstructure become comparable both interfaces have similar curvature, ultimately leading to a transition to a wetting state. This also explains why hierarchical micropatterning helps making superhydrophobic substrates more robust sustaining non-wetting states\cite{Verho10210}.

This mechanism also explains the experimentally observed phenomenology of evaporating drops\cite{XRays}, namely that non-wetting shrinking evaporating drops ultimately transition to wetting states at sizes not much larger than the microstructure scale. Furthermore, failure is triggered by deppining of the interface underneath the drop, that in contact with the microstructure, where as seen the interface has the highest curvatures and where the pinning angles can reach some depinning condition, i.e. either the contact angle with the wall of the micro-features, or a stability limit




%

\end{document}